\documentclass[%
 reprint,
 superscriptaddress,
showpacs,
 amsmath,amssymb,
 aps,
]{revtex4-1}

\usepackage{graphicx}% Include figure files
\usepackage{dcolumn}% Align table columns on decimal point
\usepackage{bm}% bold math
\usepackage{xcolor}
\usepackage{siunitx}
\allowdisplaybreaks

\usepackage{mathrsfs}

\usepackage[normalem]{ulem}

\def\lyle#1{#1} 	% updates
\begin{document}

\title{\lyle{An algebraic approach to spike-time neural codes in the hippocampus}}

\author{Federico W.~Pasini}
\thanks{These two authors contributed equally}
\affiliation{Department of Mathematics, Western University \\ London, Ontario, Canada N6A 5B7}
\affiliation{Western Academy for Advanced Research, Western University, London, ON, Canada}
\author{Alexandra N.~Busch}
\thanks{These two authors contributed equally}
\affiliation{Department of Mathematics, Western University \\ London, Ontario, Canada N6A 5B7}
\affiliation{Western Academy for Advanced Research, Western University, London, ON, Canada}
\author{J\'an Min\'a{\v c}}
\affiliation{Department of Mathematics, Western University \\ London, Ontario, Canada N6A 5B7}
\affiliation{Western Academy for Advanced Research, Western University, London, ON, Canada}
\author{Krishnan Padmanabhan}
\affiliation{Department of Neuroscience, University of Rochester Medical Center \\ Rochester, New York USA 14642}
\author{Lyle Muller}
\affiliation{Department of Mathematics, Western University \\ London, Ontario, Canada N6A 5B7}
\affiliation{Western Academy for Advanced Research, Western University, London, ON, Canada}
\date{\today}

\begin{abstract}
\lyle{Although temporal coding through spike-time patterns has long been of interest in neuroscience, the specific structures that could be useful for spike-time codes remain highly unclear. Here, we introduce a new analytical approach, using techniques from discrete mathematics, to study spike-time codes. As an initial example, we focus on the phenomenon of ``phase precession'' in the rodent hippocampus. During navigation and learning on a physical track, specific cells in a rodent's brain form a highly structured pattern relative to the oscillation of population activity in this region. Studies of phase precession largely focus on its role in precisely ordering spike times for synaptic plasticity, as the role of phase precession in memory formation is well established. Comparatively less attention has been paid to the fact that phase precession represents one of the best candidates for a spike-time neural code. The precise nature of this code remains an open question. Here, we derive an analytical expression for an operator mapping points in physical space to complex-valued spikes by representing individual spike times as complex numbers. The properties of this operator make explicit a specific relationship between past and future in spike patterns of the hippocampus. Importantly, this mathematical approach generalizes beyond the specific phenomenon studied here, providing a new technique to study the neural codes within precise spike-time sequences found during sensory coding and motor behavior. We then introduce a novel spike-based decoding algorithm, based on this operator, that successfully decodes a simulated animal's trajectory using only the animal's initial position and a pattern of spike times. This decoder is robust to noise in spike times and works on a timescale almost an order of magnitude shorter than typically used with decoders that work on average firing rate. These results illustrate the utility of a discrete approach, based on the structure and symmetries in spike patterns across finite sets of cells, to provide insight into the structure and function of neural systems.}

\end{abstract}

\maketitle

\section{Introduction}

\lyle{The brain encodes sensory information, makes decisions, and generates motor outputs through patterns of activity across large populations of neurons. It remains unknown, however, whether these patterns are made up of precisely coordinated and meaningful spike times \cite{Mainen95}, or whether the timing of the spikes is random and only their average rate is meaningful \cite{Shadlen94,London10}. Recent experimental results in the songbird singing \cite{Chi01, Srivastava17, Okubo15, Daliparthi19} and the motor system \cite{Sober18} indicate that precise spike timing can dramatically influence behavior \cite{Ahissar92, Ince10, Kara00, Tiesinga08}. In this work, we introduce an approach to study specific, individual spike patterns, using methods from discrete mathematics. As a first demonstration, we consider one of the clearest experimental examples of a spike-time pattern observed in the brain -- phase precession in the rodent hippocampus.}

\subsection{\lyle{Phase precession}}

In the rodent hippocampus during navigation on a linear track, the timing of single neuron spikes exhibits a precise relationship with the large-scale population rhythm in this brain area \cite{OKeefe93,Skaggs96}. Neurons in the CA1 region of the hippocampus fire spikes when the animal is in a specific location of an environment, creating a neural representation of that location \cite{OKeefe71,OKeefe76,Chockanathan22}. The spiking regions of individual cells, called their ``place field'', tile the environment (Fig.\,1, top). During navigation, population activity in the hippocampus oscillates at a rhythm of 8 Hz, which is termed the $\theta$-rhythm \cite{Vanderwolf69,Vanderwolf83,Bland86,Buzsaki02}. Place field sizes are ordered across the dorsal-ventral axis of the CA1 region, with cells closer to the ventral pole having progressively larger place fields \cite{Jung94,Maurer05}. As a result, the brain assembles a map of physical space along an axis of cortical space, using the theta oscillation as a metronome to tie these two representations together. The computational consequences of both this mapping and the precise spike timing relative to the $\theta$-rhythm, however, remain an open question in neuroscience.

In studying the relationship of action potential timing to the $\theta-$rhythm, researchers discovered that a single neuron begins spiking at the peak of the population $\theta$-rhythm, with the timing of spikes occurring earlier and earlier in following cycles \cite{OKeefe93}. This phenomenon is called ``phase precession'' because spikes start at a specific phase of $\theta$ and then progress to earlier and earlier phases while the animal traverses the place field (Fig.\,1, middle). This process is critical to formation of memory traces \cite{Mehta02,Robbe06,Robbe09}, which are then thought to be transferred to neocortex for consolidation and long-term storage \cite{McClelland95,Klinzing19}. The spike-time pattern formed by a population of cells exhibiting phase precession (Fig. 1, middle) is central to this process of memory trace formation, as the pattern is thought to compress behavioral sequences that last for several seconds onto the timescale relevant for synaptic plasticity \cite{Skaggs96,Mehta02}.

% ---------------------------------------------------------------- %
\begin{figure}[h]
\includegraphics[width=\columnwidth]{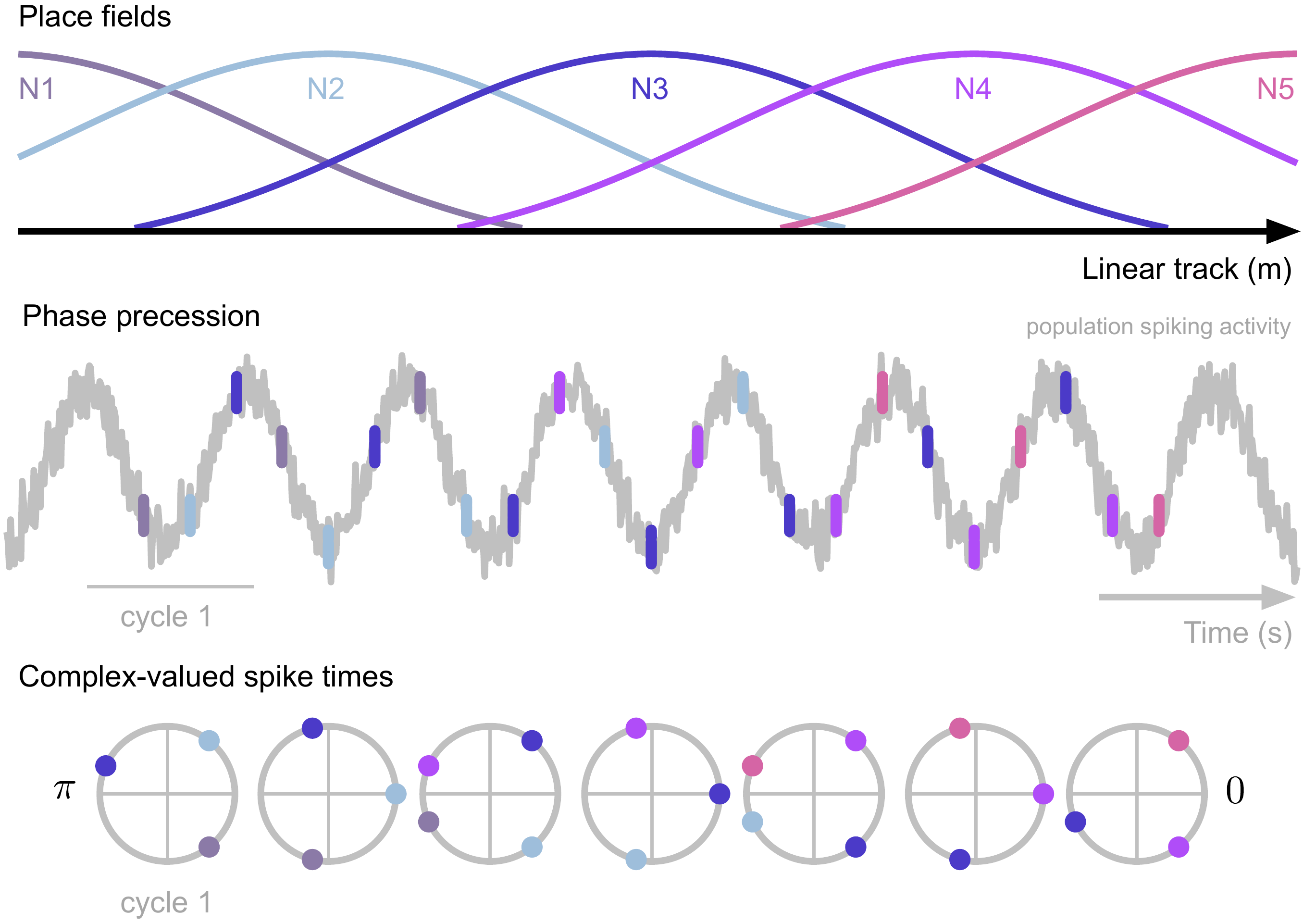}
\caption{\label{Fig1_Explanation}
Phase precession in the rodent hippocampus. (top) Depicted are five neurons (N1-N5) with active place fields during navigation from left to right on a linear track. (middle) While the rodent runs through the place field of cell N3 (blue), the phase of its spikes systematically advances with respect to the population $\theta$-rhythm (n.b.\,opposite in sign to the local field potential, relative to which phase is defined), starting close to $\pi$ at the beginning of the place field, advancing to $0$ at the place field center, and ending near $-\pi$ at the end of the place field. (bottom) A complex-valued representation of the pattern formed by the phase precession of multiple neurons N1-N5 with place field centers spaced across the track creates a compressed sequence of spike times in each individual $\theta$-cycle.}
\end{figure}
% ---------------------------------------------------------------- %

Phase precession and the compression of temporal sequences of neural activity have been well described in the dorsal pole of CA1 \cite{Skaggs96}, which represents the finest spatial scales \cite{Jung94,Maurer05}. Absent, however, is a description of what this sequence structure means for the global pattern of hippocampal activity across multiple dorsoventral levels. In other words, what can this experimental example tell us about spike-time codes across a whole brain region? The population $\theta$-rhythm itself, which was long thought to be synchronous throughout hippocampal CA1, has also recently been found to be systematically organized as a wave traveling from the dorsal to the ventral pole of CA1 \cite{Lubenov09,Patel12}. The combination of phase precession at multiple levels of the dorsoventral axis, along which the size of place fields increases linearly \cite{Kjelstrup08}, with the wave-like organization of the $\theta$-rhythm along this same axis raises the possibility that a sophisticated structure is apparent in the global pattern of spike times across the hippocampus. Understanding such a global structure in spike times could provide insight into how the hippocampal neural code is organized during the process of memory trace formation.

\lyle{In the following sections, we derive equations describing spike times in the population of neurons across the hippocampus during this phenomenon.} By representing spike phases relative to the $\theta$-rhythm in terms of complex numbers (Fig.\,1, bottom), we arrive at an operator expression relating physical space to spike times in the hippocampus. We show that this operator leads to a symmetry between past and future spike patterns in hippocampal populations, and that this symmetry reflects a specific trajectory in space and time. Further, this symmetry provides a specific meaning to the recent observation that the $\theta$-rhythm is a wave traveling across the dorsoventral axis \cite{Lubenov09}. Based on our operator expression, we then introduce a spike-based decoder that can correctly predict the animal's location. This decoder requires only the animal's starting position and spike times, and it operates on a timescale almost an order of magnitude shorter than current decoders that work on average firing rate \cite{Zhang98,Glaser20}. Importantly, this decoder, which is derived from the operator expression, allows us to relax key simplifying assumptions made in developing the mathematical approach. Taken together, these results provide fundamental new insight into mathematical approaches to spike-time codes, in addition to the specific temporal code exhibited during phase precession in the hippocampus.

\section{Results}

\subsection{\lyle{Analytical approach}} We start by introducing our notation for the hippocampal spike pattern. Experimental observations show that place field length varies along the dorsoventral axis and is constant on cross-sections of the axis \cite{Jung94,Maurer05}. In order to parametrize these quantities, we introduce the variable $\ell\in [0,1]$ to represent position along the dorsoventral axis, with $\ell=0$ being the dorsal pole. To good approximation, place field length increases linearly along the dorsoventral axis \cite{Kjelstrup08}:

\begin{equation}\label{eq:PF length}
L_\ell = L_0 + (L_1-L_0)\ell\,,
\end{equation}
where place field lengths range from less than one meter at the dorsal pole ($L_0$) to approximately 10 meters at the ventral pole ($L_1$) \cite{Kjelstrup08}.

The total phase precession of a place cell during a single traversal of its place field spans approximately a full $\theta$ cycle \cite{Skaggs96}. More precisely, the spikes of a cell systematically advance their phase with respect to the $\theta$ oscillation, with a total phase gain approaching $2\pi$. It is well documented that the phase of a spike (or the mean phase of a spike burst) of an active place cell within a theta cycle reflects the fraction of the cell’s place field the rodent has traversed at the moment of the spike \cite{OKeefe93,Skaggs96}. Note that the spike could represent a single action potential or the centroid for a burst of spikes, as typically considered in studies of phase precession \cite{Skaggs96,Mehta02}. Here we construct a model which makes the relationship between the animal's position within place fields and spike phases precise. Taking $\phi \in [-\pi,\pi]$, the spike happens at a phase $\phi$ such that the expression
\begin{equation}\label{eq:position-phase relation}
\frac{-\phi+\pi}{2\pi}
\end{equation}
\noindent equals the fraction of the place field covered at the time of the spike. Setting the space coordinate of the place field centre to be $c$, the spike phase can then be retrieved as 
\begin{equation}\label{eq:phase-position relation}
\phi = -2\pi\frac{x-c}{L_\ell}.
\end{equation}
Importantly, while the phase offset of $\theta$ will change linearly with $\ell$ as the $\theta$ wave travels over the dorsoventral axis, reaching $\pi$ at the ventral pole \cite{Patel12}, here we represent each spike's phase with respect to the {\it local} $\theta$-rhythm, instead of referencing all spikes to the phase of the $\theta$ traveling wave at the dorsal pole.

A spike happening at the physical space-time coordinates $(x_s,t_s)$ can thus be mapped to a complex valued spike phasor using \eqref{eq:PF length} and \eqref{eq:phase-position relation}. The mapping from physical space and time to complex-valued spikes can be seen as an operator $H: \mathbb{R}^2 \to \mathbb{C}$:
\begin{equation}\label{eq:operator}
H:(x_s,t_s)\to \exp\left(-2\pi i\frac{x_s-c}{L_0+(L_1-L_0)\ell}\right).
\end{equation}
As expected from experimental evidence \cite{Mehta02}, only spatial position appears in the $H$ operator. This closed-form analytical expression now allows us to understand the functional significance of phase precession in terms of symmetries in this discrete pattern of interleaved spikes in the hippocampus.

%% ------------------------------------------------------%%
\begin{figure*}[t]
\includegraphics[width=17.8cm]{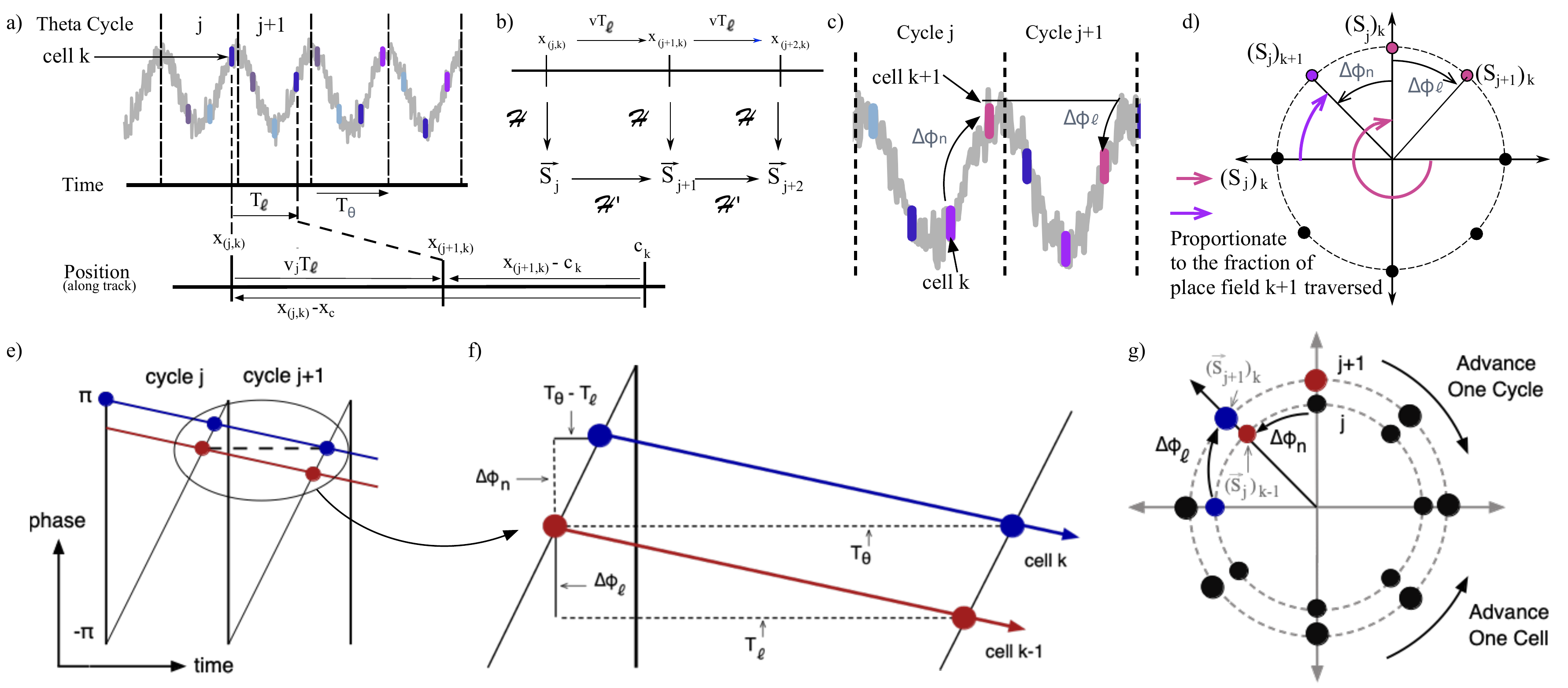}
\caption{\label{Fig2} \lyle{Phase precession schematic with computation of $\Delta \phi_\ell$ and $\Delta \phi_n$. (a) Schematic depicting the relationship between phase, time and space in the hippocampus. In theta cycle $j$, cell $k$ spikes near $-\pi$ phase, indicating place field $k$ began in the recent past. The time between the spikes of cell $k$ in $\theta$ cycles $j$ and $j+1$ is given by $T_\ell$, which is slightly shorter than the length of the $\theta$ cycle, $T_\theta$. In that time, the rat travels a distance of $v_j T_\ell$ along the track. The place field center $c_k$ and the rat's physical positions at the time of the spikes are denoted. (b) This diagram demonstrates the relationships between the vector operations for the population of neurons under consideration, where $\mathcal{H}$ maps a physical position along the track to the spikes of the full population using the shift between cells, and $\mathcal{H'}$ propagates the spike pattern of the population into the next $\theta$ cycle using the shift in time. (c) Depicts the difference between $\Delta \phi_n$ and $\Delta \phi_\ell$. (d) When these two quantities are equal, the spike pattern of the population is invariant. Advancing forwards one $\theta$ cycle simply rotates the labels of the spikes clockwise by one spike. (e) The x and y axes represent time and spike phase, respectively. The vertical black bars demarcate theta cycles. The diagonal black lines represent the phase of the $\theta$-cycle. The blue and red circles represent the spike phases of two cells at the same location along the dorsoventral axis. These phases are determined by the intersection points of the $\theta$-cycle phase (diagonal black lines) with the coloured lines, the slope of which is determined by the spiking frequency of the place cell. f)  Computation of $\Delta \phi_n$. A close-up view shows the invariant case when the phase difference between the spikes of two cells with consecutive place fields, $\Delta \phi_n$, is exactly opposite to the phase difference between spikes of the same cell in consecutive theta cycles $\Delta \phi_\ell$. The diagonal black lines have slope $2\pi / T_\theta$ by construction, since they represent the phase of the $\theta$ oscillation across by one cycle. This means $\frac{2\pi}{T_\theta} = \frac{\Delta\phi_n}{T_\theta - T_\ell}$. (g) Here, the same values are represented on the unit circle in the complex plane. Rotating clockwise by $\Delta \phi_\ell$ propagates the spike of a cell into the subsequent theta cycle. A rotation of $\Delta \phi_n$ counterclockwise is a shift forwards in space:\,it transforms the spike of cell $k$ into the spike of cell $k+1$, which has the next place field.}}
\end{figure*}

%% ------------------------------------------------------%%

\subsection{\lyle{Space-time symmetries in the $H$-operator}} We now (1) describe the time evolution of the spike pattern in terms of the $H$-operator, (2) define operations equivalent to shifting the spike pattern across $\theta$-cycles in time or across place cells in space, and then (3) use these operations to study symmetries in this space-time representation in the hippocampus. While we focus in this section on the well-studied case of rodents navigating on a linear track, the analytical form for the operator $H$ allows us to generalize quite naturally to spike patterns in two dimensions, as we will show later.

First, we describe the time evolution of the spike pattern in terms of the $H$-operator. If a cell's total phase precession spans a full $2\pi$, then it oscillates with precisely one more cycle than the $\theta$-rhythm during place field traversal. Assuming that in the $j$-th $\theta$-cycle the animal's velocity $v_j$ is constant (without requiring constant velocity on a longer timescale), the spiking frequency $f_\ell$ of a place cell at dorsoventral location $\ell$ is:

\begin{equation}\label{eq:freq}
f_\ell=f_\theta + \frac{v_j}{L_\ell}=f_\theta + \frac{v_j}{L_0 + (L_1-L_0)\ell}.
\end{equation}

Note that \eqref{eq:freq} defines the local slope of this relationship without requiring global information about the full phase precession.

\lyle{Second, we define an operation equivalent to shifting the spike pattern across $\theta$-cycles in time.} Let $c_k$ be the position of the place field center of cell $k$, and let $x_{(j,k)}$ denote the position at which cell $k$ fires during $\theta$-cycle $j$. The phase corresponding to this spike (or burst) is given by $(\vec{S}_j)_k = H(x_{(j,k)}) = \exp(-2\pi i \frac{x_{(j,k)} - c_k}{L_\ell})$. If we let $T_\ell = 1/f_\ell$ denote the time between consecutive spikes of the same cell, then the spike phase of the cell $(\vec{S}_{j+1})_k$ in the subsequent theta cycle can be described in terms of $(\vec{S}_j)_k$:

\begin{align} \nonumber
(\vec{S}_{j+1})_k &= \exp\left(-2\pi i \frac{x_{(j+1,k)} -c_k}{L_\ell}\right) \\ \nonumber
&= \exp\left(-2\pi i\frac{x_{(j,k)} + v_jT_\ell - c_k}{L_\ell} \right) \\ \nonumber
&= \exp\left(-2\pi i\frac{x_{(j,k)} -c_k}{L_\ell} \right)\exp\left(-2\pi i\frac{v_jT_\ell}{L_\ell} \right)\\ \nonumber
&= H\left(x_{(j,k)}\right)\exp\left(-2\pi i\frac{v_j}{L_\ell f_\ell} \right) \\ 
&= (\vec{S}_j)_k\exp\left(-2\pi i\frac{v_j}{L_\ell f_\ell} \right) \label{eq:lattice_relation}
\end{align}

This equation indicates that, to advance $\vec{S}_j$ to the next $\theta$-cycle, we rotate clockwise by an angle of $2\pi \frac{v_j}{L_\ell f_\ell}$.

\lyle{Third, we now specify and study the symmetry in the phase precession spike pattern. We fix} the dorsoventral location $\ell$ so that all place fields under consideration have length $L_\ell$. In order to illustrate this idea, we now consider a simplified scenario in which $v$ is constant during a single run and a population of neurons in which a new cell starts firing at phase $\pi$ on each $\theta$-cycle. In this case (Fig.~\ref{Fig2}c-g), the phase difference $\Delta \phi_\ell=-2\pi \frac{v}{L_\ell f_\ell}$ between spikes of the same cell in two consecutive $\theta$-cycles equals the difference in phase between the spikes in the same $\theta$-cycle of two cells with consecutive place field centers:

\begin{align}
\nonumber
\Delta \phi_n &=  \frac{2\pi}{T_\theta}(T_\theta - T_\ell) = 2\pi \left(1-\frac{f_\theta}{f_\ell}\right) \\
&= 2\pi\left( 1 - \frac{L_\ell f_\theta}{v}\right)\,.
\end{align}

Therefore, the following rotation transforms the spike of cell $k-1$ in $\theta$-cycle $j$ into the spike of cell $k$ in $\theta$-cycle $j$ (Fig.\,2b): 

\begin{equation}
(\vec{S}_j)_{k} = (\vec{S}_j)_{k-1} \exp\left(2\pi i \left(1 - \frac{f_\theta}{f_\ell}\right) \right)\,.\label{eq:lattice_relation2}
\end{equation} 

Equations \ref{eq:lattice_relation} and \ref{eq:lattice_relation2} allow us to derive an explicit expression for the symmetry in the hippocampal spike code. Take $\tau = (f_\ell - f_\theta)$ to be the temporal frequency of the phase precession phenomenon itself (with units \si{rad/s}) and $\chi = 1/L_\ell$ to be the spatial frequency determined by the length of a place field (with units \si{rad/m}). When $v = \tau / \chi$, the discrete spike pattern is invariant in time, i.e.\,$(\vec{S}_{j})_{k-1} = (\vec{S}_{j+1})_k~\forall\,k \in [1,N_A]$, where $N_A$ is the number of actively spiking cells \lyle{(see Methods - Complex-Valued Spikes)}. Here, the quantity $\tau / \chi$ has an important physical meaning: this invariant, with units of \si{m/s}, represents a fixed trajectory linking past to future locations on a specific space-time scale. Further, the temporal frequency determining this space-time scale is $\tau = (f_\ell - f_\theta)$, representing two key phenomena internal to the hippocampus, the $\theta$-rhythm itself and the spiking frequency of a place cell during phase precession. Finally, while we focused on the case of constant speed, the spike patterns resulting from an arbitrary time-varying movement profile (reflecting changes in speed) involve a straightforward extension of this calculation.

\begin{figure}[t]
\includegraphics[width=.95\columnwidth]{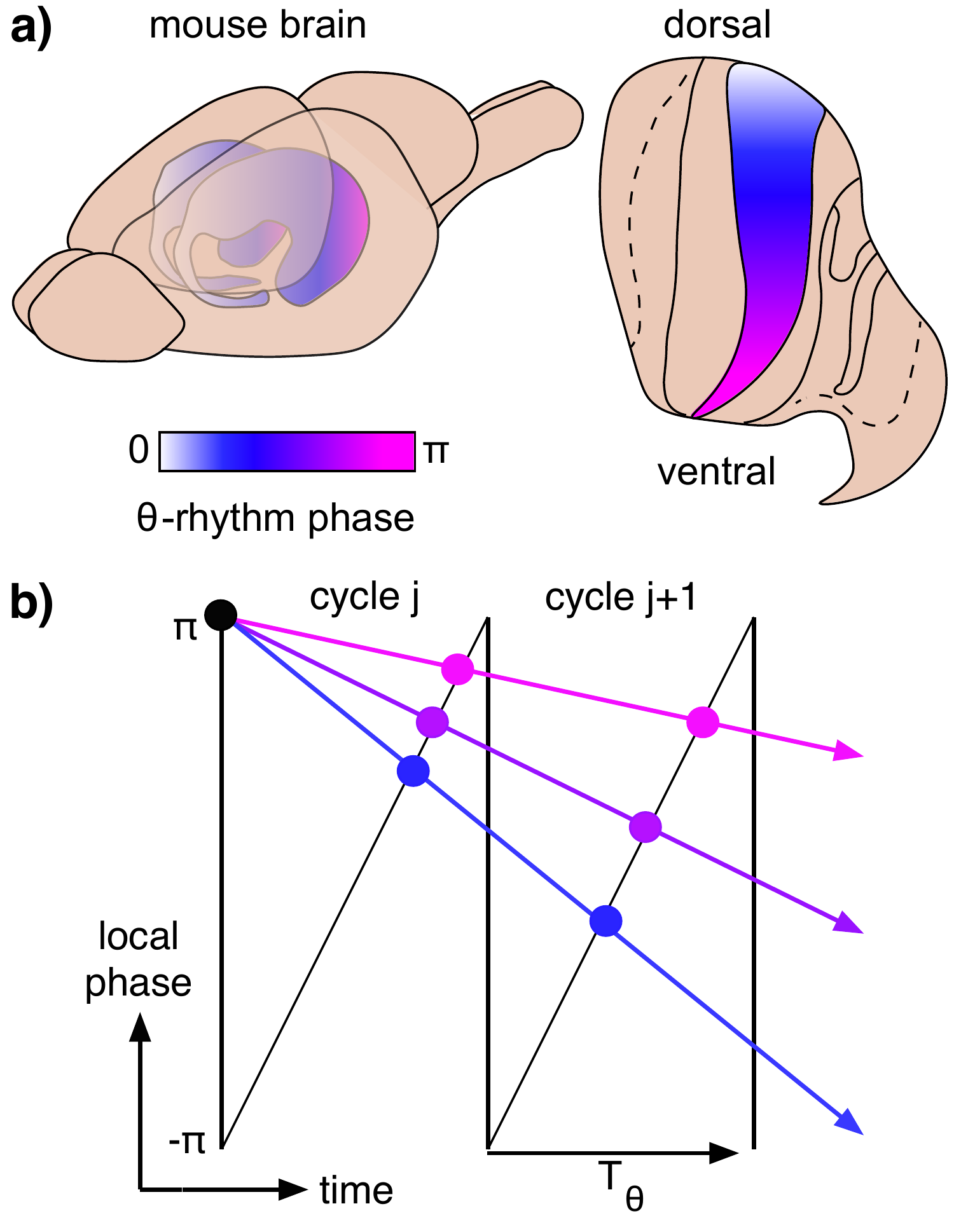}
\caption{\label{Fig3} Relationship between dorsoventral location and phase precession. (a) A mouse brain depiction is coloured by the phase offset of the $\theta$-rhythm along the dorso-ventral axis, along which the $\theta$-rhythm travels as a wave with total phase offset $\pi$ \cite{Patel12}. (b) Phase precession is depicted for three cells at different locations along the dorsoventral axis using the same colour scheme.}
\end{figure}

This mathematical approach becomes even more revealing when considering this invariant at multiple dorsoventral levels in the hippocampus. The quantity $\tau/\chi$ is independent of $\ell$. This quantity defines a space-time scale projecting a trajectory into the future. This specific space-time scale remains invariant whether the local populations represent the smallest spatial scales (``dorsal'', Fig.~\ref{Fig3}a right) or the largest spatial scales (``ventral'', Fig.~\ref{Fig3}a right). As the $\theta$-wave sweeps across CA1 during each oscillation cycle (color scale, Fig.~\ref{Fig3}a), the {\it local} populations across the dorsoventral axis represent a trajectory with a fixed scale within the context of a larger and larger cognitive map. The global organization of the hippocampal spike code during an individual $\theta$-cycle (Fig. 3b) thus represents a single, invariant trajectory linking the past and future at increasing spatial scales.
% ---------------------------------------------------------------- %
% ---------------------------------------------------------------- %

\subsection{\lyle{Spike-based decoder}} 
\begin{figure}[t]
    \centering
    \includegraphics[width=\columnwidth]{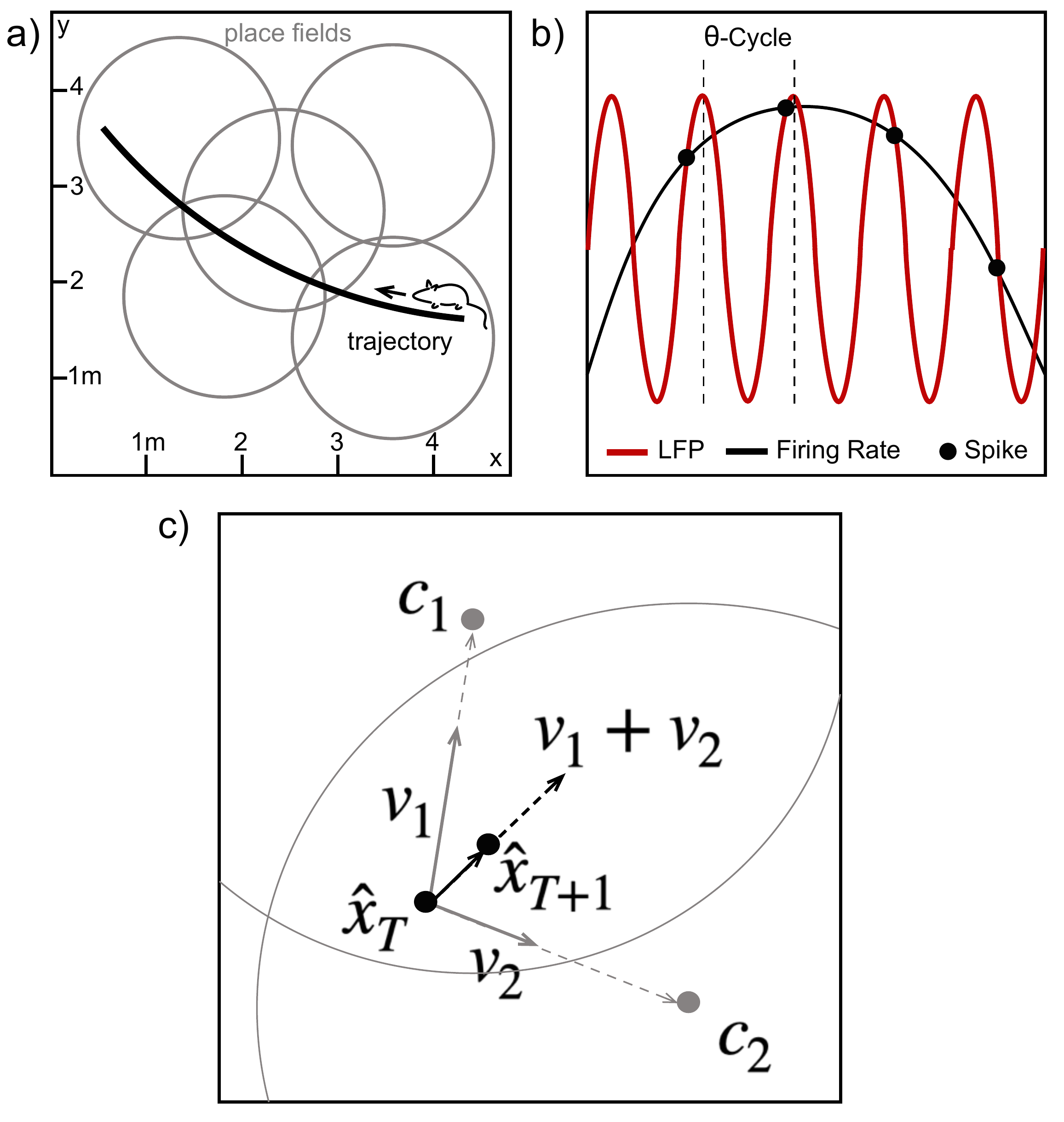}
    \caption{\label{Fig4}\lyle{Encoding and decoding spike phases: a) A 2D space tiled with place fields. Each neuron is represented by a 2D Gaussian defining the firing rate of the cell at each position in the place field. The values of the Gaussian for a particular cell along the trajectory define the black 1D curve in b). The intersections of the LFP (red) with this curve define the phase of a spike (or spike burst). These intersections occur on the rising phase before the peak firing rate (which occurs at the center of the place field) and on the falling phase after the peak. c) Depicts the decoding step to update the position estimate $\hat{x}_T$ at time $T$ for a simple case with only two place fields.}}
\end{figure}

We next sought to understand how this spatio-temporal pattern of spiking could be decoded under realistic conditions. To do this, we developed a computational approach to estimate the inverse of the $H$-operator (see Methods - Spike-Based Decoding Algorithm). \lyle{Trajectories of varying lengths and curvature were randomly generated within a 2D physical environment tiled by Gaussian place fields (Fig.~\ref{Fig4}a). A simplified model was used to generate phase precession along these 2D trajectories (Fig.~\ref{Fig4}b; see Methods - Spike-Based Decoding Algorithm). Note that neither constant speed along the trajectories nor full 2$\pi$ phase precession was assumed. For simplicity, we model uniformly sized place cells at each dorsoventral location; we note, however, that generating place fields with varying lengths has no significant effect on these results. Given the set of generated spikes, the rodent’s position is estimated by computing a weighted sum of vectors determined by the spike phases of cells with active place fields. More precisely, in each theta cycle, the set of vectors from the animal’s current estimated position to the center of each active place field is computed. The direction of each vector is given by the sign of the spike phase of the corresponding cell, depending on whether the animal is approaching or leaving the place field center. The weight of each vector in the sum is given by the fraction of the place field traversed since the previous estimate (Fig.~\ref{Fig4}c, see equation ~\eqref{eq:decoder} for fraction and normalizing factor). The sum of all contributing vectors is added to the animal’s current estimated position to produce a position update. Error improves with increasing number of active neurons, until reaching an asymptote around 30 cm at approximately 100 neurons. We start the process at an initial spatial location for the animal, apply the decoding step to update the spatial location, and then repeat this step to iteratively update the estimate of the animal’s spatial location. Critically, this spike-based decoding algorithm receives only the initial spatial location, after which it computes each new estimate based on the previous estimate. In this way, the decoded trajectories presented here are based only on an initial condition and the internally generated spike train, in much the same way the process would be implemented in the brain. Note that the error measure plotted here thus requires the decoder to estimate the correct space-time prediction, as the position decoded from spikes on a single theta cycle is compared to the subject's actual position at that same time.}

This decoding algorithm allows us to study how the spike train structure described by the $H$-operator may encode the position of the animal under realistic conditions, including biological noise in spike phase, varying speed of the animal, and phase precession in two dimensions. We first applied the decoding algorithm to linear trajectories within a 2D environment (Fig.~\ref{Fig5}a, solid line and dots), which were well handled by the algorithm. Further, the general form of this operator provides a straightforward generalization to decoding two-dimensional trajectories with arbitrary curvature (Fig.~\ref{Fig5}b, black solid line and black dots). The trajectories decoded by the algorithm are robust to noise in the spike phase (Fig.~\ref{Fig5}b, blue dots, and Fig.~\ref{Fig5}d). In contrast, random spike phases resulted in decoded trajectories extending far from the correct path and in different directions on each iteration (Fig. \ref{Fig5}b, red dots). These results show that a decoding algorithm inspired by the $H$-operator introduced in this work can faithfully decode spatial locations from the temporal pattern of spikes alone, and on the timescale of a single $\theta$-cycle.

\section{Discussion}

\lyle{In this work,} we have introduced a discrete mathematical approach to spike patterns that are a key component of memory formation \cite{OKeefe93,Mehta02}. A central idea in neuroscience is that memories are stored across large groups of neurons \cite{Sejnowski99}. Neural activity is composed of discrete spiking events, but nearly all models of memory focus on the continuous rate of spikes, rather than their discrete timing. This approximation has been convenient because precise spike times are generally considered to be stochastic; however, more and more recent experimental work suggests that precise timing of spikes across groups of cells may be critical to memory formation \cite{Mehta02,Robbe09,Jaramillo17}.

By starting with a specific experimental observation in the hippocampus, where the clearest demonstration of a precise spike-time pattern has been found, we have introduced a novel, discrete approach to the problem of spike times in memory. We first introduced a model for this precise pattern of spike times, and we then derived an analytical expression for the operator mapping points in physical space to complex-valued spike times. We find this operator leads to a clear space-time symmetry between past and future spike patterns in the hippocampus. This theoretical approach unites the role of the hippocampus in spatial navigation with its role in memory formation. Finally, we introduced a spike-based decoder that can successfully predict an animal's spatial location under realistic noisy conditions, and on the timescale of a single $\theta$-cycle ($\sim$125 ms), a timescale almost an order of magnitude shorter than typically used with rate-based decoders \cite{Zhang98,Glaser20}.

\subsection{\lyle{Difference from previous work}}

\lyle{Theoretical work on phase precession has largely focused on its role in memory formation, since these spike-time structures compress behavioural timescales onto the scale of a single $\theta$-cycle \cite{Skaggs96}. This sequence compression has been causally implicated in the process of memory formation by pharmacological manipulation of NMDA receptors \cite{Mehta02}, can allow temporal-order learning \cite{Reifenstein19}, and has been studied in both spiking network models \cite{Tsodyks96} and analytically \cite{Geisler10}. Further, \cite{Geisler10} studied how phase precession (where cells spike at a frequency just higher than the population) can be consistent with the oscillation in the population spiking activity. Finally, a position-theta-phase model is proposed in \cite{McClain19} to investigate the modulation of firing rate by running speed. These studies demonstrate interest in developing theoretical approaches to phase precession. This previous work, however, placed less emphasis on studying phase precession as a spike-time code. Due to the lack of theoretical work on this topic, many critical aspects of phase precession remain unexplored, including what behaviourally relevant features could be encoded by phase precessing populations and how phase precession generalizes beyond one-dimensional linear tracks (as considered in \citealp{Skaggs96,Mehta02}). In this work, we have utilized approaches from discrete mathematics to understand the spike-time structure involved in phase precession. Our operator expression, which relates points on a trajectory to complex-valued spikes in a hippocampal population, provides not only a formula for the spike times in the simplified one-dimensional scenario considered in the analytical approach, but also a straightforward generalization to two (or more) dimensions. Further, this operator expression reveals a symmetry in the hippocampal representation, which encodes not just the animal's position but also trajectories linking past and future positions. Finally, we show that the dorsal-ventral sweep of theta plays a specific role in this code: the travelling theta wave unites local representations of trajectory across CA1 in the context of an expanding spatial map.}

\subsection{\lyle{$\theta$ traveling wave in the hippocampus}}

\lyle{These results provide insight into the global organization of hippocampal activity during memory formation and the recent observation that $\theta$ is a wave that travels from dorsal to ventral portions of the hippocampus \cite{Lubenov09,Patel12}. The hippocampus is thought to be where a ``cognitive map'' \cite{OKeefe76,OKeefe78} emerges, such that information about an animal's location during navigation serves as the brain's ``GPS system'' \cite{McNaughton06}. If the central function of the hippocampus were only to precisely encode spatial location, we might expect the temporal organization of activity during a $\theta$ cycle to sweep from the ventral, coarsest spatial scale of representation to the dorsal, finest spatial scale, analogous to a radar sweep that would initially scan across coarse scales and progressively narrow its focus to identify an object's location. If, on the other hand, the fundamental function of the hippocampus is linking past and future, then a sweep from the dorsal to the ventral pole would represent a trajectory linking past and future positions at a fixed space-time scale within a larger and larger spatial map. Interestingly, recent experiments have reported that neural activity can encode possible future positions and that this neural activity can be precisely coordinated both within and between theta cycles \cite{Johnson07,Kay20,Joshi22}. Though these reports are not directly aligned with the phenomenon studied here, these experiments represent potential avenues for connecting this mathematical approach with analysis of neural recordings in future work. Importantly, our mathematical results crystallize the role of the $\theta$ traveling wave in the hippocampus, which is to establish memories as a link between past and future \citep{McClelland95}. This role, which is in agreement with the direction of propagation observed in experimental data \cite{Lubenov09,Patel12}, also provides one of the first clear computational roles for traveling waves in the brain \cite{Muller18}.}

\subsection{\lyle{A discrete approach to spike-time codes}}

\lyle{Spike times are often considered to be random and modelled as stochastic point processes. As such, it is not the events themselves, but rather the rates at which they occur, that are most often related to behaviour. This may be due, in part, to the lack of mathematical tools capable of accounting for the precise timing of spikes. Approaches from discrete mathematics may provide a new avenue for exploring how spike timing may contribute meaningfully to neural computation. By studying specific spike time patterns, and leveraging their algebraic properties (such as symmetries), it may be possible to uncover general theoretical principles for the role of spike timing in neural computation. In this work, we introduce an operator approach to phase precession, one of the best known examples of a specific spike-time pattern. Representing neural spikes as complex numbers allows for the exploration of symmetries, like the rotations described above, not easily observable in a linear variable. Many possibilities exist for future work to develop general methods for arbitrary spike-time codes in populations of the hippocampus and neocortex.}

\begin{figure}[t]
    \centering
    \includegraphics[width=\columnwidth]{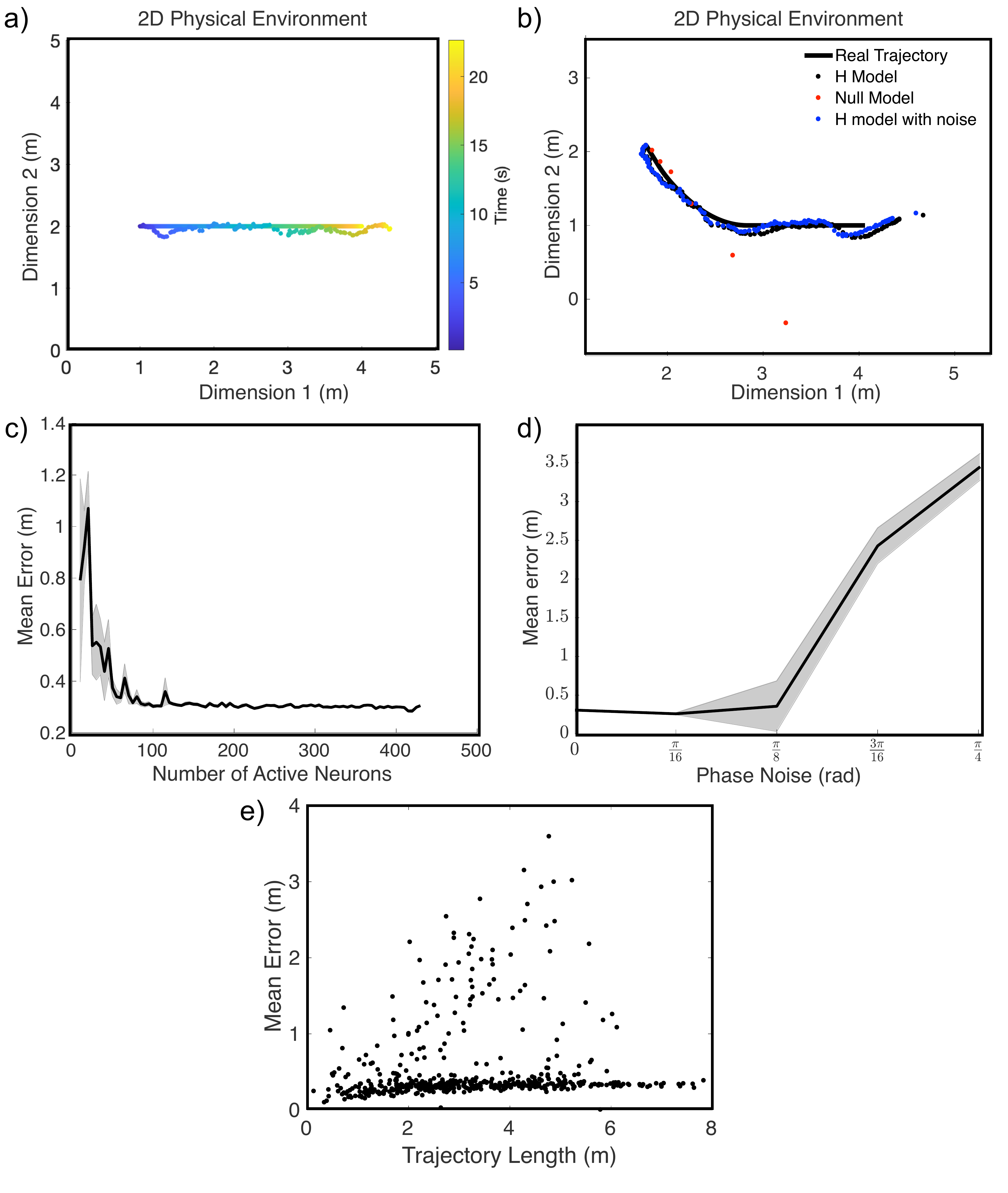}
    \caption{\label{Fig5}Spike-train decoding based on the $H$-operator. (a) 1D trajectory in 2D environment. A linear trajectory (solid line) is plotted for a simulated rodent running from left to right (color changing from blue to yellow in time), along with dots for decoded positions in each theta cycle (same color code). Simulated place cells have 1 m fields, leading to 169 active neurons on this trajectory. (b) An example 2D trajectory (solid line) is plotted along with the results from several decoding models: the $H$-model (black dots), the $H$-model with $\pi/16$ phase noise, and a null model in which each spike is assigned a random phase value. Place cells have 1 m fields, and 388 cells are active along the 3.36m trajectory. (c) For the example trajectory in b), mean error is plotted as a function of the number of place fields active along the trajectory. Different numbers of place fields were randomly generated to tile the 2D space surrounding the trajectory, resulting in different numbers of cells with place fields overlapping the trajectory. (d) Decoding error for the same trajectory is plotted as a function of added phase noise. For each value of phase noise, 10 iterations were used to produce the plot. Solid lines represent mean decoding error averaged over realizations, and shaded regions represent standard error. \lyle{(e) Mean error is plotted as a function of trajectory length.}}
    %\label{fig:decoding_analysis}
\end{figure}

\appendix\section{Methods}
\subsection{Complex-Valued Spikes}

We begin by recalling the $H$-operator and notation from the main text. For cells at a fixed location $\ell \in [0,1]$ along the dorsoventral axis of CA1, all place fields have length $L_\ell = L_0 + (L_1 - L_0)\ell$. Suppose that cell $k$ has place field center $c_k$ and spikes at the physical position $x_{(j,k)}$ in theta cycle $j$. Then the spike phase of cell $k$ in theta cycle $j$ is given by:
\begin{equation} \label{H operator}
(\vec{S}_j)_k = H(x_{(j,k)}) = \exp\left(-2 \pi i \frac{x_{(j,k)} - c_k}{L_\ell}\right)
\end{equation}
If the total phase precession of a cell during place field traversal spans a full $\theta$ cycle, the cell is active for precisely one more cycle than the $\theta$ rhythm during the traversal. Assume that in each $\theta$ cycle, $j$, the rat's velocity is constant and given by $v_j$; however, this velocity need not be constant along the whole trajectory. Then, the spiking frequency in $\theta$ cycle $j$ of a place cell at dorsal-ventral location $\ell$ can be written as: 
\begin{equation}\label{eq:freq2}
f_\ell=f_\theta + \frac{v_j}{L_\ell}
\end{equation}
%

%% address the full 2pi precession assumption: it is the slope of the line that matters, not the full 2pi part. If it doesn't start at -pi, the fraction it does start at reflects how much of the place field has already been traversed (?) also if you don't have the full 2pi, you're essentially doing a small rotational shift - shifting the line down a little, and dropping off the last set of spikes. so it still works (because it's the slope that matters) 
Note that this equation defines a local slope (Fig. \ref{Fig2}e,f, red and blue lines), and does not require global information about the full phase precession (i.e., it still holds when the cell does not precess a full $2\pi$). 
Let $T_\ell = 1/f_\ell$ denote the time between spikes of the same cell in two consecutive theta cycles. Then, the physical distance travelled between those two spikes is $v_jT_\ell$, as in Fig. ~\ref{Fig2}a. This quantity can be used to define the shift required to propagate the spike phase of cell $k$ from theta cycle $j$ into cycle $j+1$. In this way, it is unnecessary to compute $(\vec{S}_{j+1})_k$ from the $H$-operator directly: it can also be estimated from $x_{(j,k)}$, $T_\ell$, and $v_j$ as follows.  
\begin{align}
(\vec{S}_{j+1})_k &= H( x_{(j+1,k)}) \\ \nonumber
&=\exp\left(-2\pi i \frac{x_{(j+1,k)} -c_k}{L_\ell}\right) \\ \nonumber
&= \exp\left(-2\pi i\frac{x_{(j,k)} + v_jT_\ell - c_k}{L_\ell} \right) \\ \nonumber
&= \exp\left(-2\pi i\frac{x_{(j,k)} -c_k}{L_\ell} \right)\exp\left(-2\pi i\frac{v_jT_\ell}{L_\ell} \right)\\ \nonumber
&= H\left(x_{(j,k)}\right)\exp\left(-2\pi i\frac{v_j}{L_\ell f_\ell} \right) \\ \nonumber
&= (\vec{S}_j)_k\exp\left(-2\pi i\frac{v_j}{L_\ell f_\ell} \right)
\end{align}
Therefore, rotating clockwise by an angle of $2\pi \frac{v_j}{L_\ell f_\ell}$ propagates the spike of cell $k$ forward to the next theta cycle. 

Now, let $v_j = v$ be constant throughout the trajectory, and suppose a new neuron begins spiking every theta cycle. In this case, the phase difference $\Delta \phi_\ell=-2\pi \frac{v}{L_\ell f_\ell}$ between the spikes of the same cell in two consecutive theta cycles equals the difference in phase between the spikes in the same theta cycle of two cells with consecutive place field centers. This quantity is given by $\Delta \phi_n$ and can be computed from Fig. ~\ref{Fig2}f as: $\Delta \phi_n =  \frac{2\pi}{T_\theta}(T_\theta - T_\ell) =2\pi (1-\frac{f_\theta}{f_\ell}) = 2\pi( 1 - \frac{L_\ell f_\theta}{v})$. Here, the positive slope represents moving from the spike of the red cell to that of the blue cell. Since the red cell has precessed more than the blue cell, we know the place field of the red cell began before that of the blue cell; i.e., if blue denotes cell $k$, then red denotes cell $k-1$. 

We can therefore define the following rotation to transform the spike of cell $k-1$ in theta cycle $j$ into the spike of cell $k$ in theta cycle $j$:
\begin{equation}(\vec{S}_j)_k = (\vec{S}_j)_{k-1} \exp\left(2\pi i \left(1 - \frac{f_\theta}{f_\ell}\right) \right)\end{equation}
Equivalently, $ (\vec{S}_j)_{k-1} = (\vec{S}_j)_{k} \exp(-2\pi i (1 - \frac{f_\theta}{f_\ell}) )$, shifts backwards in space, moving from one cell to another with a place field that began earlier in the trajectory. 

As in Fig. \ref{Fig2}g, we define the shift from one cell to the next by the rotation of $2\pi(1 - \frac{f_\theta}{f_\ell})$ counter clockwise. Importantly, this phase difference, $\Delta \phi_n$, is exactly opposite $\Delta \phi_\ell$. We can therefore shift forwards in time (i.e.~propagate the spike of a cell forwards one theta cycle) by rotating clockwise, and shift forwards in space (i.e.~shift to the cell with the subsequent place field) by rotating counter clockwise. Furthermore, when considering a population of $N_A$ active cells at dorsal-ventral location $\ell$, in which a new cell begins spiking every $\theta$ cycle, the phase pattern of the population remains invariant across the trajectory. In this case, the spike phases of the neurons are equally spaced around the unit circle in the complex plane, and advancing forwards in time simply rotates the neuron indices from spike to spike. These rotations can be used to define the following vector operations, depicted in Fig. \ref{Fig2}b, which consider the aforementioned population of cells as a whole. 

First, map the animal's position along the track to the population spike pattern in one $\theta$ cycle using the operation: $\mathcal{H}: \mathbb{R} \to \mathbb{C}^{N_A}$ maps $x_{(j,1)} \to \vec{S_j}$, with
\begin{equation} 
(\vec{S}_j)_{k} = (\vec{S}_j)_1 \exp\left(2\pi i k\left(1 -\frac{f_\theta}{f_\ell}\right) \right)
\end{equation}
Here, cell $1$ is the active cell with the earliest place field center; i.e, in theta cycle $j$, cell $1$ reaches the end of its place field and spikes close to phase $-\pi$. This operation maps the physical location of a single spike to the spike pattern of the full population.

Second, map the spike pattern in one theta cycle to the spike pattern in the next theta cycle using the shift between $\theta$ cycles: $\mathcal{H'}: \mathbb{C}^{N_A} \to \mathbb{C}^{N_A}$ maps  $\vec{S}_j \to \vec{S}_{j+1}$ with
\begin{equation}
\vec{S}_{j+1} = \vec{S}_j \exp\left(-2\pi i\frac{v}{L_\ell f_\ell} \right)
\end{equation}
These vector operations can be composed, as illustrated in Fig. \ref{Fig2}b. As such, in the simplified case of constant speed, it is unnecessary to recompute the population spike pattern of the $N_A$ active cells individually in each theta cycle using the H operator. Instead, the $\mathcal{H}$ operator can be used to define the population spike pattern from one physical position, and then the $\mathcal{H'}$ operator can be used to propagate the spike pattern forwards in time. In this case, the spike pattern is invariant in time; i.e., $(S_{j+1})_{k+1} = (S_j)_k$ for all cells $k$ and $\theta$ cycles $j$, \lyle{as is made clear by re-expressing $(S_{j+1})_{k+1}$ as two specific rotations of $(S_j)_k$:}

\begin{align}
\nonumber
(S_{j+1})_{k+1} &= (S_{j+1})_k\exp\left(2\pi i\left( 1 - \frac{f_\theta}{f_\ell}\right) \right)\\ \nonumber
& \qquad\text{ (shift forward one cell)} \\ \nonumber
&= (S_j)_k\exp\!\left(-2\pi i \frac{v}{L_\ell f_\ell}\right)\exp\!\left(2\pi i\!\left( 1 - \frac{f_\theta}{f_\ell}\right)\! \right)\\ \nonumber
&\qquad\text{(shift forward one theta cycle) } \\ \nonumber
&= (S_j)_k \text{ precisely when }\\
&-2\pi i \frac{v}{L_\ell f_\ell} + 2\pi i\left( 1 - \frac{f_\theta}{f_\ell}\right) = 0
\end{align}

In this case, we have that: $$\frac{v}{L_\ell f_\ell} = 1 - \frac{f_\theta}{f_\ell}
\implies \frac{v}{L_\ell}= f_\ell - f_\theta 
\implies v = (f_\ell - f_\theta) L_\ell\,,$$
which yields the relationship $v = \tau / \chi$ where $\tau = (f_\ell - f_\theta)$ is the difference between the cellular frequency and the $\theta$ frequency, and $\chi = 1/L_\ell$ is the spatial frequency determined by the length of the place fields. 

\subsection{Spike-Based Decoding Algorithm}

To implement the decoder, we first generate a trajectory in a two-dimensional space tiled with $N_C$ place fields. Place field centers $\{c_k\}_{k=1}^{N_C}$ are randomly generated and uniformly distributed across the space. The place fields all have the same diameter $L$ reflecting cells in a specific dorsal-ventral location in the hippocampus. The trajectory is given by the positions $\{x_j\}$ corresponding to $\theta$-cycles $j \in [1, N_T]$. Note that the time-resolution of the generated trajectory is much finer than the length of a $theta$-cycle. Within each cycle, the simulated animal travels over a short line segment, which we represent by a set of points, $x_j$. The decoded position, $\hat{x}_j$, is a point representing the decoder's estimate for the position of set $x_j$.

 %Since spikes of different cells within the same theta cycle happen at slightly different times and phases, the physical positions they refer to are also slightly different (in accordance with $\eqref{eq:position-phase relation}$ and $\eqref{eq:phase-position relation}$. Consequently, the calculated positions $x_j$ based on each single active cell form a discrete sequence rather than identifying a single point. Since the time-resolution of the generated trajectory is much finer than the length of a $\theta$-cycle, for each $\theta$-cycle j, with good approximation $x_j$ is a short line segment. However, the resulting decoded position, $\widehat{x}_j$, is a point. %Note that the time-resolution of the generated trajectory is much finer than the length of a $\theta$-cycle, so for each $\theta$-cycle $j$, $x_j$ is a line segment. However, the resulting decoded position, $\hat{x}_j$, is a point. 

Since the sequence of line segments $\{x_j\}$ captures both position and time, there is no need to assume the animal moves at a constant speed. As such, we implement trajectories with varying speeds, accelerations, lengths, and curvatures. We used a simple model to generate phase precession along the trajectory using a Gaussian place field and sinusoidal theta rhythm. The combination of the firing rate profile and the theta oscillation determines at which phase the spike occurs in each $\theta$-cycle. Following previous work \cite{Mehta02}, the moment that the sinusoidal oscillation crosses the level of excitatory input determined by the place field at each spatial location determines the spike. This moment is on the rising phase for the first half of the place field and on the falling phase for the second. Note that the spike could represent a single action potential or the centroid for a burst of spikes, as typically considered in studies of phase precession \cite{Skaggs96,Mehta02}.

This method is used to generate a matrix of spikes, $S$, where entry $S_{(j,k)} \in [-\pi, \pi]$ gives the spike phase of cell $k$ during $\theta$-cycle $j$. Note that, while $N_C$ represents the total number of place fields tiling the space, only a small fraction of them are active during any given $\theta$-cycle. The number of cells which are active in a specific theta cycle, $N_{A,j}$, is given by the number of place fields which overlap the line segment $x_j$. If cell $k$ does not spike during $\theta$-cycle $j$, $S_{(j,k)} = NaN$.   

The decoder takes as input the initial position, $x_1$, and the matrix of spikes $S$. Decoding proceeds as follows:

% vector sum for decoding 

For $j = 1$, $\hat{x}_j = x_{j,1}$, the first point in the line segment $x_1$. For $j \in [2,N_T]$, where $N_T$ is the total number of $\theta$-cycles in the trajectory, 
\begin{equation}\label{eq:decoder}
\hat{x}_j =  \hat{x}_{j-1} + \frac{ \pi }{ N_{A,j} } \sum_{k = 1}^{N_C} F_{j,k} \frac{(-1)^\alpha (\hat{x}_{j-1} - c_k)}
{ \|(-1)^\alpha ( \hat{x}_{j-1} - c_k) \| } 
\end{equation}
where \[ \alpha = \begin{cases} 
      0 & S_{(j,k)} > 0 \\
      1 & S_{(j,k)}\leq 0 
   \end{cases}
\]  
and $F_{j,k}$ represents the fraction of place field $k$ traversed in theta cycle $j$. This fraction is determined by: $$F_{j,k} =\frac{L}{2\pi} ( S_{(j,k)} - S_{(j-1,k)} + \pi).$$ 

For each $\theta$-cycle, the vectors $(\hat{x}_{j-1} - c_k)$ are computed for each active cell $k$. Each of these vectors forms a line segment from the animal's current estimated position, $\hat{x}_{j-1}$, to the center of a place field corresponding to an active cell. The directions of these vectors are determined by the exponent $\alpha$; i.e., the animal moves towards the center of a place field with positive spike phase, and away from the center if the spike phase is negative. Note that here spike phases are defined with respect to the local field potential, which is opposite to the population oscillation discussed in the main text. The vectors are then normalized to unit length and multiplied by $F_{j,k}$, so the magnitude of the vector is determined by the fraction of the place field traversed during that $\theta$-cycle. The sum of these vectors estimates the animal's movement during that $\theta$-cycle. However, it overestimates the amount of distance covered by the animal due to overlap between place fields. The normalizing factor $\frac{\pi}{N_{A,j}}$ was chosen experimentally to reduce the effects of this overlap. Finally, this vector sum is added to the animal's current position to produce an estimated position update.   

To compute the accuracy of the decoded trajectory, we consider the minimum Euclidean distance from each estimate $\hat{x_j}$ to the line segment $x_j$. Define the line segment $x_j := \{x_{j,i}\}_{i=1}^{N_{S,j}}$, where $N_{S,j}$ denotes the number of points in the line segment, determined by the speed of the animal's movement during $\theta$-cycle $j$. Then, the accuracy of the estimate for a single theta cycle is given by: 
$$d_j := \underset{i}{\min} \| x_{j,i} - \hat{x}_j \|$$
For the entire trajectory, we then define the mean error $e_m := \frac{1}{N_T} \sum_{j=1}^{N_T} d_j$ and cumulative error $e_c := \sum_{j=1}^{N_T} d_j$.

In Fig. ~\ref{Fig5}d, the mean error is plotted for one example trajectory with varying amounts of active place fields and added phase noise. In Fig. ~\ref{Fig5}e, he mean error is plotted for 500 random trajectories of varying lengths. To produce this plot, the number of place fields tiling the space was held constant, and 500 randomly generated trajectories were decoded in separate realizations of the decoding algorithm. Note that trajectories near the boundary of the space result in increased error due to edge effects, which are straightforward to address but not considered here. The example in \ref{Fig5}b was chosen to avoid the boundaries, but the trajectories in \ref{Fig5}d were completely random.

% ================================================================ %
% ================================================================ %
%                                                                  %
% ACKNOWLEDGEMENTS                                                 %
%                                                                  %
% ================================================================ %
% ================================================================ %

\begin{acknowledgements}
This work was supported by BrainsCAN at Western University through the Canada First Research Excellence Fund (CFREF), the NSF through a NeuroNex award (\#2015276), the Natural Sciences and Engineering Research Council of Canada (NSERC) grant R0370A01, SPIRITS 2020 of Kyoto University, Compute Ontario (computeontario.ca), Compute Canada (computecanada.ca), and the Western Academy for Advanced Research. J.M.~gratefully acknowledges the Western University Faculty of Science Distinguished Professorship in 2020-2021. A.B.~gratefully acknowledges support from the BrainsCAN Graduate Studentship Program. K.P. was supported by NSF CAREER (1749772), NIMH (R01MH11392), the Schmitt Foundation, and the Cystinosis Research Foundation.
\end{acknowledgements}
% ================================================================ %
% ================================================================ %
%                                                                  %
% REFERENCES                                                       %
%                                                                  %
% ================================================================ %
% ================================================================ %

\bibliography{references.bib}

\end{document}